# Room Temperature Spin Pumping in Topological Insulator $Bi_2Se_3$

Mahdi Jamali,[1] Joon Sue Lee,[2] Yang Lv,[1] Zhengyang Zhao,[1] Nitin Samarth[2] and Jian-Ping Wang[1*]

[1]Department of Electrical and Computer Engineering, University of Minnesota, MN 55455

[2]Department of Physics, The Pennsylvania State University, University Park, PA 16802

**Three-dimensional (3D) topological insulators are known for their strong spin-orbit coupling and the existence of spin-textured topological surface states which could be potentially exploited for spintronics. Here, we investigate spin pumping from a metallic ferromagnet (CoFeB) into a 3D topological insulator ($Bi_2Se_3$) and demonstrate successful spin injection from CoFeB into $Bi_2Se_3$ and the direct detection of the electromotive force generated by the inverse spin Hal effect (ISHE) at room temperature. The spin pumping, driven by the magnetization dynamics of the metallic ferromagnet, introduces a spin current into the topological insulator layer, resulting in a broadening of the ferromagnetic resonance (FMR) linewidth. We find that the FMR linewidth more than quintuples, the spin mixing conductance can be as large as $3.4 \times 10^{20} m^{-2}$ and the spin Hall angle can be as large as 0.23 in the $Bi_2Se_3$ layer.**

[*]Electronic address: jpwang@umn.edu

Spintronics is a rapidly growing field of study in past two decades. It has found several applications in memory[1,2], logic[3] and sensing devices[4,5] that utilize both the spin and charge of electrons. Spintronics also is a field of surface and interface science and engineering where astounding effects including giant magnetoresistance[6] and tunneling magnetoresistance[7] occur. One of the main challenges of the spintronic devices is scattering of the electron spin due to the impedance-mismatch between the magnetic and nonmagnetic layers[8,9]. Recently the spin pumping effect has been investigated intensively due to its genuine freedom from impedance-mismatch issue[10,11]. In a spin pumping device, spin current is injected from the magnetic layer into the nonmagnetic layer due to the precessional magnetization dynamics[12–14]. Moreover, spin pumping is a means by which several important dynamical properties of the magnetic and nonmagnetic layers can be quantified including spin mismatch conductance and spin Hall angle[15].

Topological insulators (TIs) are insulating in the bulk like ordinary insulators but have gapless edge/surface states [16,17]. These states are possible due to the strong spin-orbit interaction and fundamental symmetries of the materials[18]. In addition, these states are highly robust and are protected against impurities or defects by the time-reversal symmetry[19]. The presence of topologically protected helical surface states has been verified by means of angle-resolved photoemission spectroscopy[20–22] and scanning tunneling microscopy[23–25]. The TI surface electrons exhibit massless Dirac particles like graphene. Moreover, their spins are locked at right-angle to their momentum[19,22,26], which means all the surface carriers that move in the same direction are intrinsically 100% spin-polarized. This aspect of TI materials is very attractive for spintronic applications. Since spintronic devices are highly influenced by the behavior of spin at interfaces, TI materials are of natural interest in this context.

Recent theoretical proposals have focused on the behavior of spin-polarized carriers on the surface of TIs and the possibility of large spin-orbit torques using a magnetic layer[27,28]. Although it is expected that TI materials show a large spin Hall angle at room temperature due to their very strong spin orbit coupling, recent experimental results based on spin pumping have reported a very weak signal at room temperature[29,30]. In contrast, measurements of permalloy/$Bi_2Se_3$ bilayers using spin torque FMR (ST-FMR) reveal a very large figure of merit characterized by a spin torque angle between 2 – 3.5 even at room temperature [31]. In this study, we carry out a complementary experiment to ST-FMR and demonstrate the injection of a large spin current into a 3D TI via the spin pumping effect at room temperature. We further find that the spin injection efficiency deduced from the broadening of the FMR linewidth and spin mismatch conductance can fluctuate significantly from sample to sample. Our experimental results suggest that the spin Hall angle can vary by more than an order of magnitude from sample to sample of $Bi_2Se_3$ TIs.

In our study, we use $Bi_2Se_3$ as the TI since this is a material that has been extensively studied both theoretically and experimentally. The existence of topological surface states at room temperature is well established in this material [32–37]. Electrical and optical measurements report that bulk of $Bi_2Se_3$ is a semiconductor with a moderately large band gap of about 0.35 eV[35,38–40]. Depending on the $Bi_2Se_3$ sample preparation conditions, the charge transport properties can vary significantly with a tendency toward an n-type extrinsic semiconductor behavior due to the presence of vacancies or dislocations and also because of excessive selenium[41,42]. The $Bi_2Se_3$ thin films used in this study are synthesized on semi-insulating InP(111)A substrates by molecular beam epitaxy (MBE) (see Methods for the detailed procedure). A typical atomic force microscopy (AFM) image of a $Bi_2Se_3$ film surface is given in Fig. 1(a) showing 1-nm steps of quintuple layers (QLs). Each QL consists of five atomic layers of Se-Bi-Se-Bi-Se as can be seen in Fig. 1(b). The temperature dependence of the resistivity of the $Bi_2Se_3$ samples shows metallic behavior characteristic of a degenerately doped semiconductor [Fig. 1(c)]. This is consistent with the resistivity of about 1000 µΩ.cm at room temperature that indicates a high level of doping of the Bi$_2$Se$_3$ film. Hall measurements of the patterned Bi$_2$Se$_3$ films [illustrated in supplementary Fig. S2(a)] reveal a 2D carrier density of a 10 QL film of about 2.5 x 10$^{13}$ cm$^{-2}$, indicating that the chemical potential is above the bottom of the bulk conduction band.

We investigate the spin dynamics at the Bi$_2$Se$_3$/CoFeB interface using spin pumping (see Methods for fabrication details). The sample geometry is shown in Fig. 2(a) and consists of a thin ferromagnetic layer placed on top of the topological insulator film. The magnetization dynamics is excited in the magnetic layer using a GHz-rf field. Due to the spin pumping effect, a net spin current ($J_s$) is injected into the TI layer. The large spin-orbit coupling of the TI converts this spin current into a charge current $J_c = \theta_{SHE}(2e/\hbar)J_s$ that can be read out as $V_{ISHE}$ where $\theta_{SHE}$ is the spin Hall angle. The actual fabricated device with the measurement setup is shown in Fig. 2(b). The magnetization dynamics is excited using an asymmetric coplanar waveguide in the GS form utilizing a microwave sinusoidal source. The width of the signal (S) and ground (G) lines are 60 μm and 180 μm respectively and spacing between them is 60 μm. The coplanar waveguide is isolated from the magnetic and TI layers by MgO (50 nm) oxide layer. For the magnetic layer, CoFeB (5 nm) is deposited on the Bi$_2$Se$_3$ film using a Shamrock sputtering system. The surface of the TI is slightly etched using an *in-situ* argon ion miller to provide a fresh interface for the magnetic layer. Using a subsequent optical lithography process, the TI layer is patterned into a rectangular shape of 1500 μm × 620 μm. In this report, we present the experimental results for four samples BS1 (10 QL), BS2 (10 QL), BS3 (10 QL), and BS4 (5 QL).

The output dc-voltage is induced by ISHE which originates in skew scattering created by the spin-orbit interaction. The electromotive force induced by the ISHE is given by[22] $E_{ISHE} \propto J_S \times \sigma$ where $J_S$ is the spin current injected from CoFeB in TI ($\hat{z}$) and $\sigma$ is the spin polarization vector of the spin current ($\hat{y}$).

Fig. 3(a) shows the output dc-voltage due to the ISHE for an excitation frequency of 4 GHz and an excitation amplitude of 2.5 V for the sample BS1(10 QL). The output signal polarity alters as the direction of the magnetic field reverses while the absolute value of the resonance field remains the same value of 10 mT which is consistent with typical spin pumping spectra[12]. In addition, by increasing the input excitation frequency to 6 and 8 GHz (Figs. 3(b) and (c)), the resonance field shifts to 21.2 and 36.7 mT, respectively [see supplementary Fig. S3 for contour plots]. These results are consistent with an FMR spectrum governed by the Kittel formula $f = \frac{\gamma}{2\pi}\sqrt{H(H+M_s)}$ where $M_s$ is the saturation magnetization of the CoFeB and $\gamma$ is the gyromagnetic ratio[12]. The spin pumping signal has the form of a symmetric Lorentzian function. Usually there is a contribution from the anisotropic magnetoresistance (AMR) and/or the anomalous Hall effect (AHE) of the magnetic layer (CoFeB) in the output voltage. Both AMR and AHE have the form of asymmetric Lorentzian functions and can be isolated from the output signal. By fitting the experimental spectra to the form $V = V_{SP}\frac{\Delta H^2}{\Delta H^2 + (H-H_r)^2} + V_{Asym}\frac{\Delta H(H-H_r)}{\Delta H^2 + (H-H_r)^2}$, the symmetric and asymmetric components are extracted (see Figs. 3(a)-(c)). Here, $H_r$ is the resonant field and $\Delta H$ is the linewidth of the spin pumping signal.

In Figs. 3(d)-(f) the spin pumping spectra are given for the sample BS2 (10 QL) [see supplementary Fig. S4 for the contour plot]. The results show that the linewidth of the spin pumping signal increases by more than 2 times compared to the sample BS1. The linewidth of the sample BS1 are 5.4, 6.2, and 7.7 mT, while the ones for the sample BS2 are 15.3, 17.3, 18.2 mT for the excitation frequencies of 4, 6, and 8 GHz, respectively. Moreover, the amplitude of the output voltage drops by more than an order of magnitude in sample BS2. The asymmetric component of the output signal is dissociated from the output signal as can be seen in Figs. 3(d)- (f). Since the spin pumping linewidth is associated with the nature of the damping of magnetization at the TI/CoFeB interface, we conclude that the $Bi_2Se_3$/CoFeB interfaces behave differently in sample BS1 and BS2.

In Figs. 4(a) & (b), the spin pumping spectra are given for excitation amplitudes of 1.0, 1.5, 2.0, and 2.5 V for sample BS1 and BS2, respectively. The amplitude of the spin pumping increases by increasing the excitation amplitude for both positive and negative fields. Furthermore, the broadening of the output signal for BS2 is larger than BS1 for all excitation amplitudes. In Figs. 4(c)-(d), the amplitude of the output signal peak position is drawn for different microwave excitation amplitudes. In both samples BS1 and BS2, the output amplitude increases quadratically relative to the input amplitude which is consistent with the spin pumping properties[12,43].

In Fig. 5(a), the spin pumping spectra are shown for samples BS1-BS4 at a constant excitation frequency of 6 GHz and an excitation amplitude of 2.5 V. The data clearly show a wide variation of the output signal amplitude amongst these four samples. The amplitudes are 122, 13.2, 68, and 1.5 µV for BS1-BS4, respectively. There is a slight change of the resonance field due to slight variations of the saturation magnetization of the four samples. In the inset of Fig. 5(a), the normalized spin pumping signals are plotted versus the FMR signal. The linewidth of the spin pumping signal is always larger than the FMR signal in the four samples. Moreover, BS2 has the largest value of 17.3 mT and BS3 has the smallest value of 4.5 mT while the FMR linewidth is 2.3 mT. Fig. 5(b) shows the linewidth of samples BS1-BS4 overlaid with the FMR signal linewidth under different excitation frequencies. As expected, all BS1-BS4 have a linewidth that is much larger than the FMR signal. Moreover, there is a wide variation of the linewidth among these four samples.

The broadening of the spin pumping linewidth is closely related to the spin mixing conductance at the interface of $Bi_2Se_3$/CoFeB. The spin mixing conductance can be evaluated from the relation $g^{\uparrow\downarrow} = \frac{2\sqrt{3}\pi M_s \gamma d_{CFB}}{g\mu_B \omega}(\Delta H_{SP} - \Delta H_{FMR})$ where $d_{CFB}$ is the thickness of the CoFeB, $\omega$ (= $2\pi f$) the excitation frequency, $\mu_B$ is the Bohr magneton, and $g$ is the Landau $g$-factor. In Fig. 5(c), the resonant field is given for different excitation frequencies of spin pumping in samples BS1-BS4 and for FMR. All the samples show the same behavior with a slight difference in the resonance field. By fitting the data to the Kittel formula, the saturation magnetizations have been extracted independently [see supplementary Fig. S5]. The spin mixing conductances are found to be $6.5 \times 10^{19}$, $26 \times 10^{19}$, $2.926 \times 10^{19}$, and $14 \times 10^{19}$ $m^{-2}$ for BS1-BS4, respectively, at the excitation frequency of 6 GHz.

Knowing the spin mixing conductance, the spin current density injected from the ferromagnet into the TI $Bi_2Se_3$ layer can be calculated from $j_s = \frac{g^{\uparrow\downarrow}\gamma^2 h_{rf}^2 \hbar [M_s\gamma + \sqrt{(M_s\gamma)^2 + 4\omega^2}]}{8\pi\alpha^2[(M_s\gamma)^2 + 4\omega^2]}$ [44] where $h_{rf}$ is the excitation rf-field and $\alpha$ is the Gilbert damping coefficient. The damping coefficient can be calculated from $\alpha = \frac{\sqrt{3}\gamma \Delta H}{2\omega}$ [12,45]. In order to evaluate the spin current density, we need to estimate the rf-field carefully. We use a vector network analyzer to characterize the input impedance at different excitation frequencies and calculate the $h_{rf}$, accordingly [see supplementary Fig. S6]. The spin current density given by $\frac{2e}{\hbar}j_s$ is about $1 \times 10^7$, $3.3 \times 10^6$, $1.0 \times 10^7$, and $7.6 \times 10^6 A/m^2$ for samples BS1-BS4 at the excitation frequency of 6 GHz.

The output voltage can be related to the spin current and the spin Hall angle according to the formula $V_{ISHE} = \frac{w\theta_{SHE}\lambda_N \tanh(d_N/2\lambda_N)}{d_N\sigma_N + d_F\sigma_F}(\frac{2e}{\hbar})j_s$ [44], where $d_N$ is the thickness of the $Bi_2Se_3$ and $\lambda_N$ is the spin coherence length of the TI channel. In Fig. 5(d) the spin Hall angle (SHA) has been calculated for different resonance frequencies in sample BS1-BS4 [see supplementary Fig. S7 for details]. The SHA at the excitation frequency of 6 GHz are 0.2, 0.012, 0.048, and 0.009 for samples BS1-BS4, respectively. As seen, SHAs vary by more than an order of magnitude from sample to sample. Our calculated SHA of sample BS1 is more than two orders of magnitude larger than of the recent report on the spin pumping into TI[30]. Moreover, the extracted SHA of sample BS1 is consistent with the large spin torque angle measured recently in $Bi_2Se_3$ using ST-FMR[31]. Since the current density involved during our measurement is very small, our results are free from spurious effects like Joule heating. From Fig. 5(d), we see a weak dependence of the SHA on the excitation microwave frequency. Over an excitation frequency range of 3 to 9 GHz, the SHAs obtain maximum value of 0.22, 0.015, 0.05, and 0.012 in samples BS1-BS4, respectively.

The wide fluctuation of spin pumping signals from sample to sample could be explained based on the nonuniform composition of $Bi_2Se_3$ at its interface with the magnetic layer. Recently, it is shown that surfaces of TIs are very non-uniform in terms of composition[46]. Since the $Bi_2Se_3$/CoFeB interface plays the major role in the spin injection, the large variations of the spin pumping characteristics could be associated with non-uniform $Bi_2Se_3$ surface. Moreover, the decapping process could also be responsible for modification of the TI surface (see Methods for the detailed procedure) considering Bi2Se3 has strong thermoelectric properties[47]. In addition, surface roughness can modify the spin pumping efficiency[48] . As shown in Fig. 1(a), the surface of $Bi_2Se_3$ is not very smooth and potentially it can alter the spin pumping signal.

In summary, we have carried out spin pumping experiments at $Bi_2Se_3$/CoFeB interfaces. Our results clearly demonstrate the successful pumping of a spin current into the TI layer at room temperature and a large spin Hall angle. Moreover, we find that the spin pumping characteristics vary from sample to sample, likely due to a non-uniform CoFeB/$Bi_2Se_3$ interface. The spin Hall angle changes in the range of 0.01 up to 0.22. Our experimental results suggest that 3D topological insulators are very promising for the spintronic applications due to their large spin Hall angle and FMR broadening though there are challenges to overcome non-uniform surface properties.

**Acknowledgement**

This work was partially supported by the C-SPIN center, one of six STARnet program research centers and National Science Foundation Nanoelectronics Beyond 2020 (Grant No.NSF NEB **1124831**).

**Methods**

**Device fabrication**. Bi2Se3 thin films are synthesized on semi-insulating InP(111)A substrates by molecular beam epitaxy (MBE) in ultrahigh-vacuum (UHV) systems (low $10^{-10}$ Torr). After desorbing native oxide on the InP(111)A substrate under As flux, the substrate is transferred to another UHV chamber without breaking vacuum for $Bi_2Se_3$ growth. At a substrate thermocouple temperature of 350°C, single crystalline $Bi_2Se_3$ films are grown using thermal evaporation of high-purity (99.999%) elemental Bi and Se via conventional Knudsen cells with a Se:Bi beam equivalent pressure ratio of ~15:1 and a growth rate ranging from ~0.36 nm/min to ~0.83 nm/min. The morphology of the surface is monitored by reflection high-energy electron diffraction (RHEED) during the film growth (See supplementary Fig. S1). Next, the substrate is cooled down to ambient temperature under the Se flux for minimizing defects like Se vacancies. Moreover, Se serves as the capping layer for transferring the sample from MBE into the sputtering system.

An ultra-high vacuum (UHV) six-target Shamrock sputtering system with an *in-situ* ion-milling source has been used for the ferromagnetic layer deposition. The Se capping layer is removed in an AJA sputtering system with a built-in heater. Samples are heated up to 270±20 °C for an hour and then cooled down. Samples are heated up for long enough time to confirm complete removal of the capping layer. Samples are then immediately transferred to our Shamrock system for the magnetic layer deposition. The TI surface is slightly etched using ion-miller before CoFeB deposition. $Co_{20}Fe_{60}B_{20}$ (5 nm) has been deposited for the magnetic layer and the film is capped with MgO (2 nm) to prevent the magnetic layer oxidation. Next, photolithography has been used to define TI/ferromagnet regions. After coating the substrate with a negative-tone photoresist, the resist is patterned into rectangles with a size of 1500 μm × 620 μm. The patterns are transferred to the TI/ferromagnet layer using argon ion etching and subsequent resist removal. In order to isolate the ferromagnetic structure from the coplanar waveguide, MgO (50 nm) is sputtered. After patterning the coplanar waveguides and contacts, Cr (5nm)/ Cu (100 nm) is deposited by Ebeam evaporator followed by lift-off of the resist. Samples BS1 (10 QL), BS2 (10 QL), BS3 (10 QL), and BS4 (5 QL) are from four different batch of samples prepared by MBE.

**Measurements**. The samples are characterized in a high frequency probe station with *in-situ* magnetic fields (up to 1 kOe) and using GSG probes from GGB Industries, Inc. A microwave signal generated by a 20 GHz Agilent microwave generator is injected into the asymmetric coplanar waveguide and the output dc-voltage is measured using a Keithley 2182A nanovoltmeter. During the measurement, a constant microwave frequency is applied to the waveguide and the output dc-voltage is tested at each magnetic field.

Figure Captions

FIG. 1. (a) The surface morphology of the Bi2Se3 film taken by AFM. The AFM image shows 1-nm steps of quintuple layers (QLs). (b) Each QL consists of five atomic layers of Se-Bi-Se-Bi-Se. (c) The resistivity of the grown $Bi_2Se_3$ measured at different temperatures.

FIG. 2. (a) Schematic of the device utilized for the spin pumping experiment. A thin magnetic layer is placed on top of the topological insulator. An external rf-magnetic field, excites the magnetization dynamics. Precessional magnetization dynamics injects a net spin current ($J_s$) into the adjacent TI layer. Due the strong spin-orbit coupling of the TI layer, this spin current is translated into a charge current that can be characterized as the output inverse spin Hall voltage. (b) An optical micrograph of the actual fabricated device. The magnetization dynamics is excited using an asymmetric coplanar waveguide in the GS form and a microwave source.

FIG. 3. The spin pumping measured for the sample BS1. The output dc-voltage measured at excitation frequencies of (a) 4, (b) 6, and (c) 8 GHz for various bias fields. The experiment data (black) is overlaid with the symmetric Lorentzian (red) and anti-symmetric Lorentzian (blue) curve fitting. Right column shows the spin pumping signal for the sample BS2 measured at the excitation frequency of (e) 4, (f) 6, and (g) 8 GHz. The experiment data (black) is overlaid with the symmetric Lorentzian (red) and anti-symmetric Lorentzian (blue) curve fitting.

FIG. 4. The spin pumping output signal for the excitation amplitude of 2.5, 2.0, 1.5, and 1.0 V for the sample BS1 at the excitation frequency of 6 GHz (a) and for the sample BS2 at the excitation frequency of 5 GHz (b). The peak amplitude of the spin pumping signal

at different excitation amplitude for the sample BS1 at the excitation frequency of 6 GHz (c) and for the sample BS2 at the excitation frequency of 5 GHz (d).

FIG. 5. (a) The spin pumping spectra for the sample BS1 (10 QL), BS2 (10 QL), BS3 (10 QL), and BS4 (5 QL). The microwave excitation frequency is 6 GHz and the input excitation amplitude is 2.5 V for all the measurements. In the inset the normalized spin pumping signals for the samples BS1, BS2, BS3, and BS4 are given together with the FMR spectra. (b) The linewidth of the FMR and of the spin pumping signals of the samples BS1, BS2, BS3, and BS4. (c) The resonance fields for different excitation frequencies for the FMR and for the samples BS1, BS2, BS3, and BS4. (d) Spin Hall angles are extracted for BS1-BS4 at different excitation frequencies.

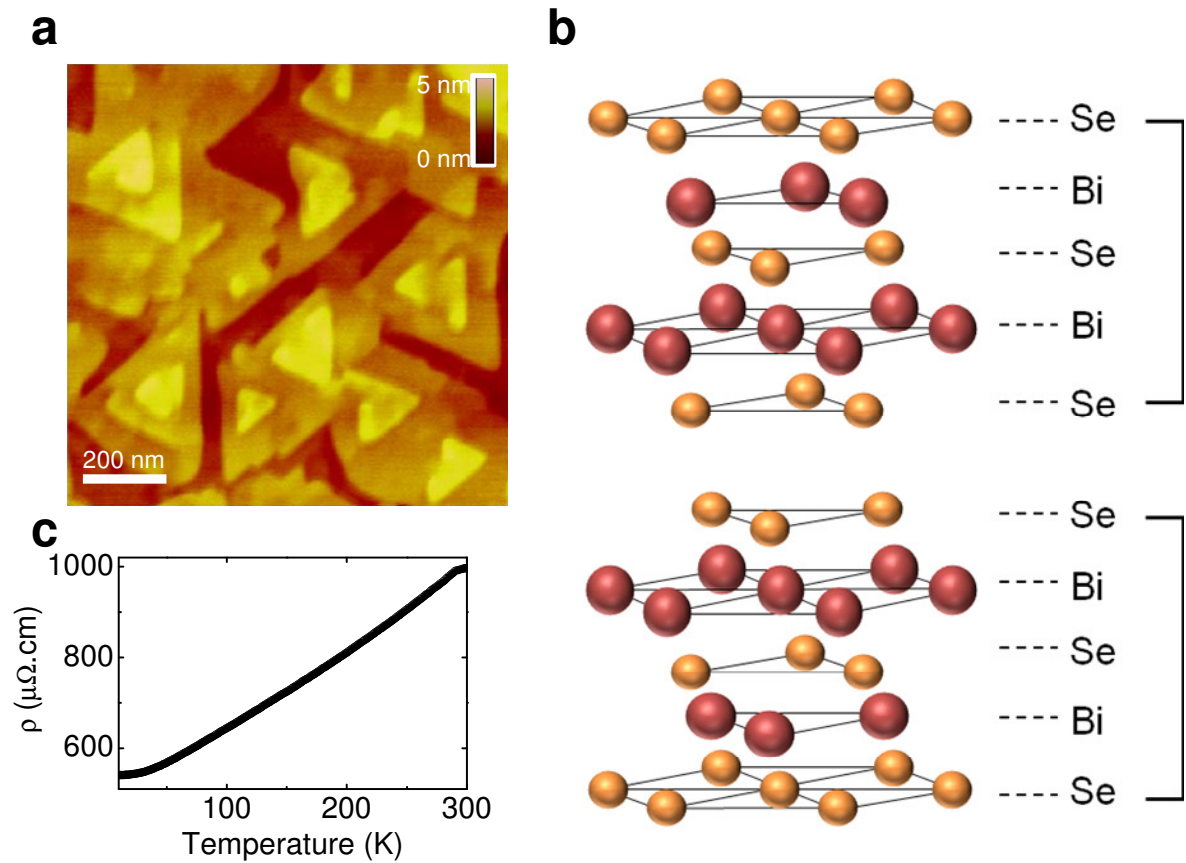

Figure 1.

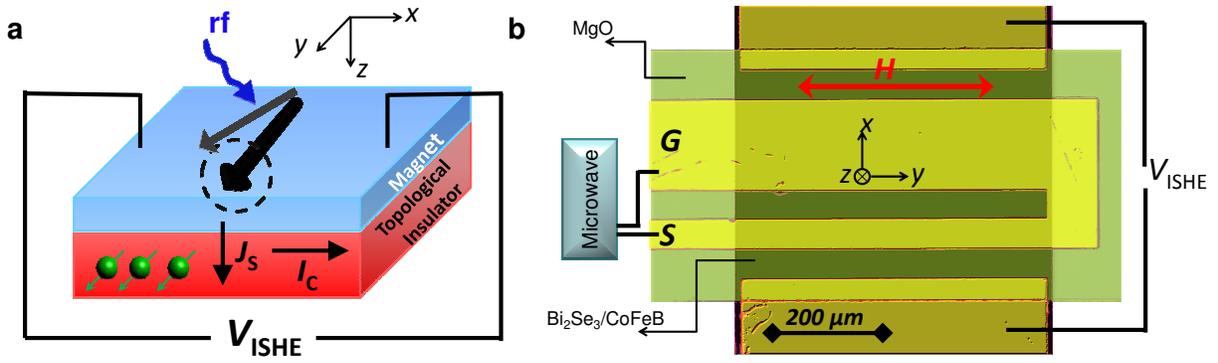

Figure 2.

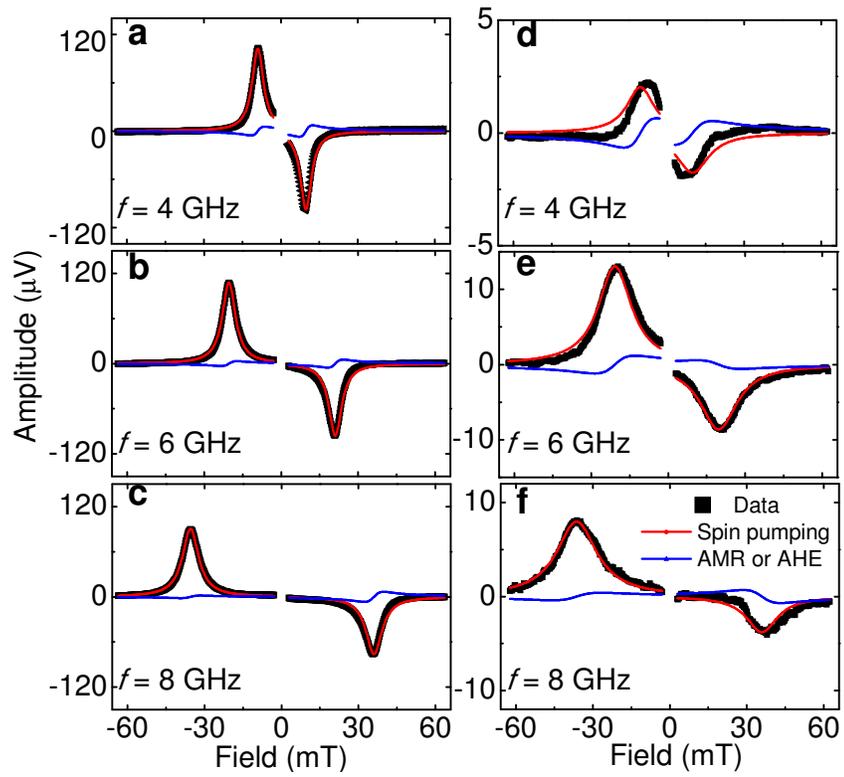

Figure 3.

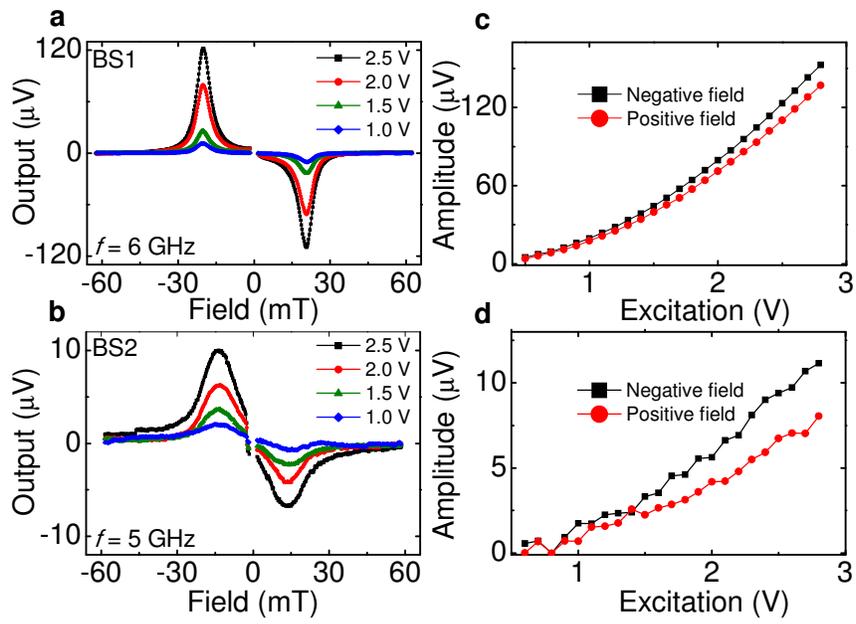

Figure 4.

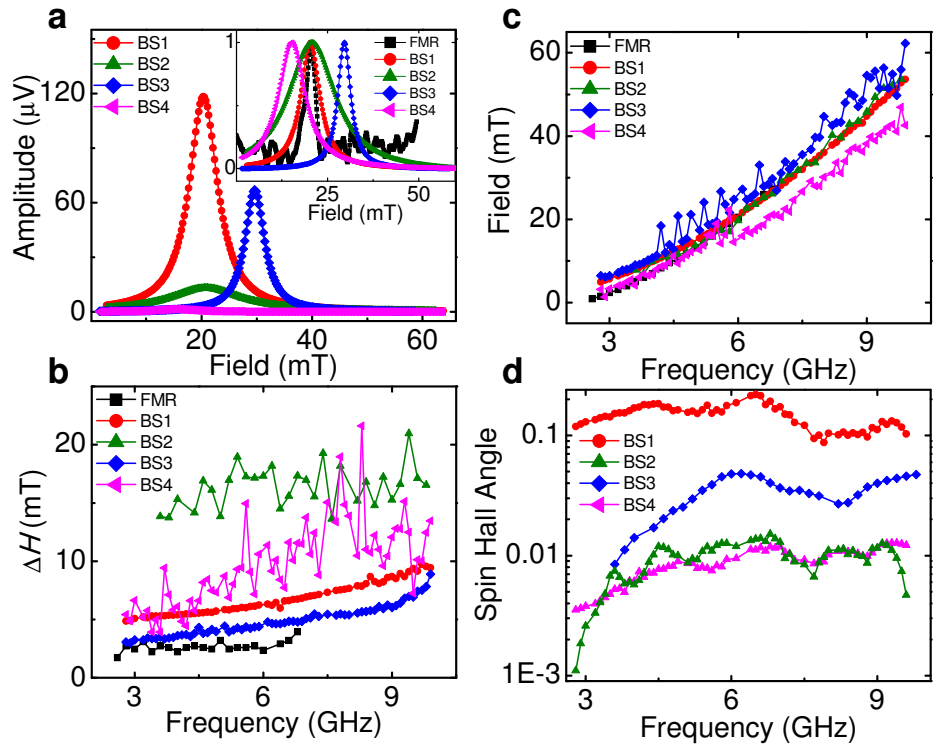

Figure 5.

# Room Temperature Spin Pumping in Topological Insulator $Bi_2Se_3$

Mahdi Jamali,[1] Joon Sue Lee,[2] Yang Lv,[1] Zhengyang Zhao,[1] Nitin Samarth[2] and Jian-Ping Wang[1*]

[1]Department of Electrical and Computer Engineering, University of Minnesota, MN 55455

[2]Department of Physics, The Pennsylvania State University, University Park, PA 16802

[*]Electronic address: jpwang@umn.edu

- **$Bi_2Se_3$ Growth and Characterization**
- **Spin Pumping and Spin Hall Angle**

    1. *Contour plot of the spin pumping spectra in sample BS1 and BS2*
    2. *Extraction of magnetic properties*
    3. *Characterization of the coplanar waveguide input impedance*
    4. *Extraction of the spin Hall angle*

- **Bi$_2$Se$_3$ Growth and Characterization**

As explained in the Methods section of the main manuscript, Bi$_2$Se$_3$ thin films are synthesized on semi-insulating InP(111)A substrates by molecular beam epitaxy (MBE) in ultrahigh-vacuum (UHV) systems (low $10^{-10}$ Torr). The morphology of the surface is monitored by reflection high-energy electron diffraction (RHEED) during the film growth. A typical RHEED pattern the Bi$_2$Se$_3$ surface is given in Fig. S1.

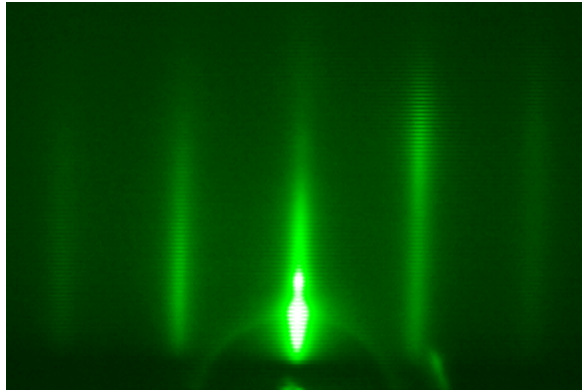

Figure S1: A typical RHEED pattern of the Bi$_2$Se$_3$ surface.

Electrical properties of the Bi$_2$Se$_3$ thin films, with Se layer decapped, are characterized using a Hall-bar geometry with a channel dimension of 650 x 400 µm$^2$, fabricated by conventional photolithography and Ar-plasma dry-etching. Hall measurements [Fig. S2(c)] reveal that the 2D carrier density of a 10 QL film is ~2.5 x $10^{13}$ cm$^{-2}$, indicating that the chemical potential is above the bottom of the bulk conduction band. The considerable contribution of charge conduction through bulk results in the metallic behavior of the temperature dependence of the longitudinal resistance, as shown in Fig. S2(d). In Fig. S2(a) and (b), the longitudinal magneto-resistance (MR) with a magnetic field perpendicular to the plane are given. One can see the quantum corrections of weak anti-localization to the classical MR near zero magnetic field at low temperature, which are seen in 3D topological insulator thin films [1,2] due to the π Berry phase of electrons in the Dirac surface state [3].

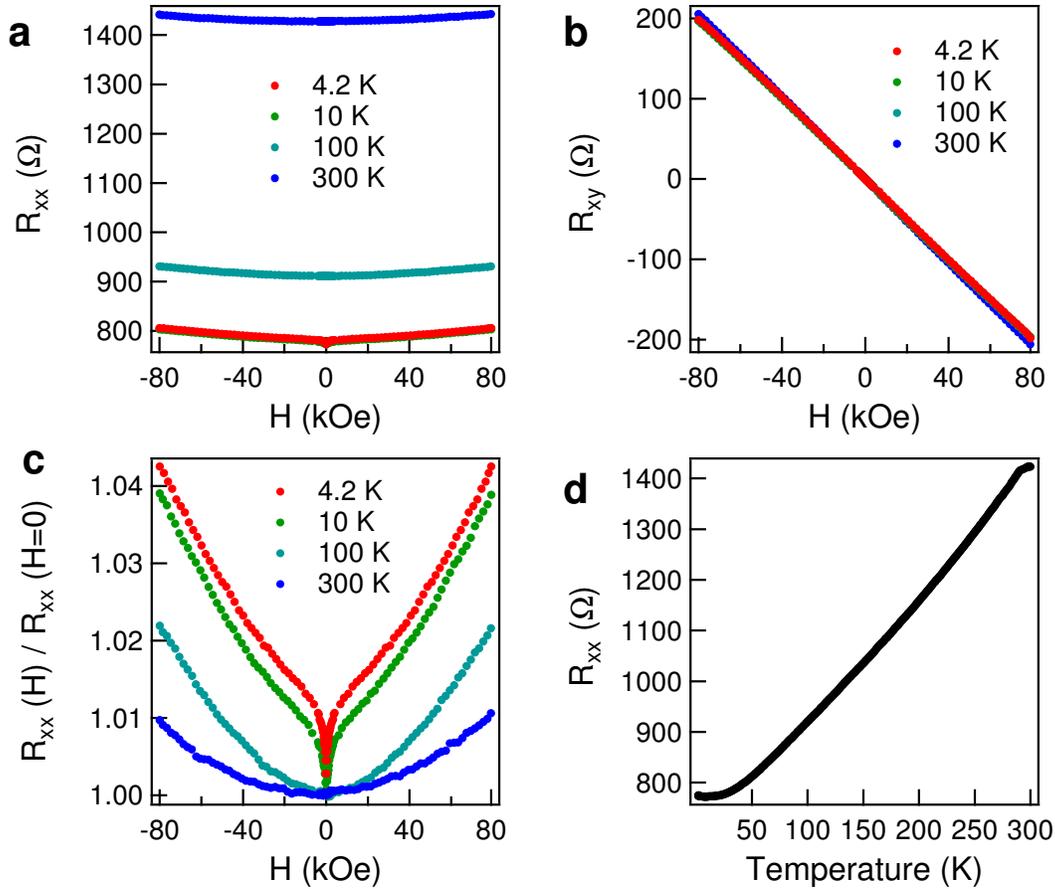

Figure S2: (a) The magnetoresistance of the patterned Bi2Se3 film measured at 4.2, 10, 100, and 300 K. (b) The Hall resistance of the patterned film tested at 4.2, 10, 100, and 300 K. (c) The normalized magnetoresistance for different temperatures. (d) The temperature dependence resistance of $Bi_2Se_3$.

- **Spin Pumping and Spin Hall Angle**

    1. *Contour plot of the spin pumping spectra in sample BS1 and BS2*

Spin pumping experiments have been performed on several $Bi_2Se_3$ samples with thickness of 5 and 10 QL. Figure S3 and S4 are the contour plot of the output voltage under different excitation frequencies and resonant fields for samples BS1 and BS2, respectively. Although both of these samples have 10 QL of $Bi_2Se_3$, their output spectra are very different in terms of amplitude and linewidth.

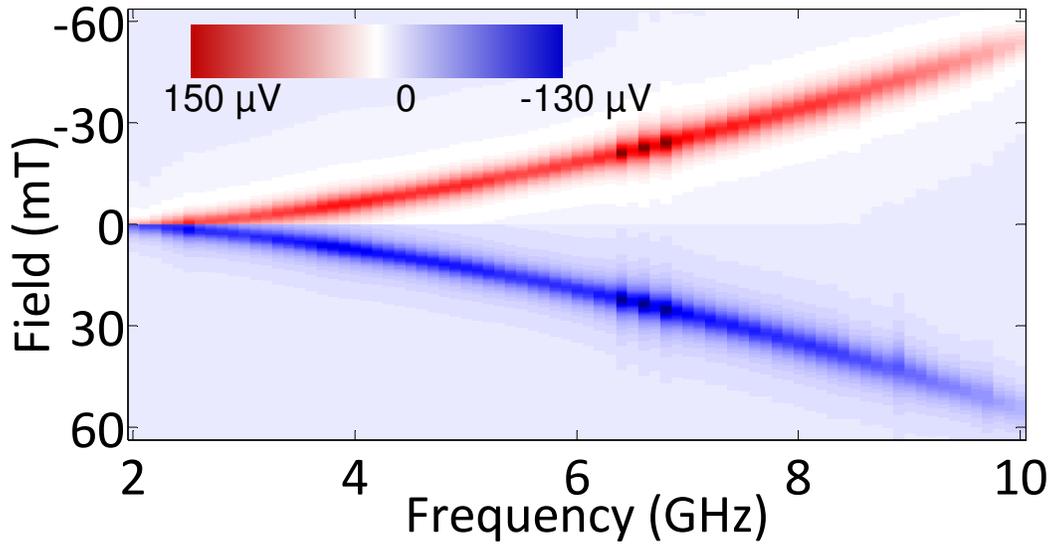

Figure S3: The contour image of the spin pumping spectra at various excitation frequencies for Sample BS1.

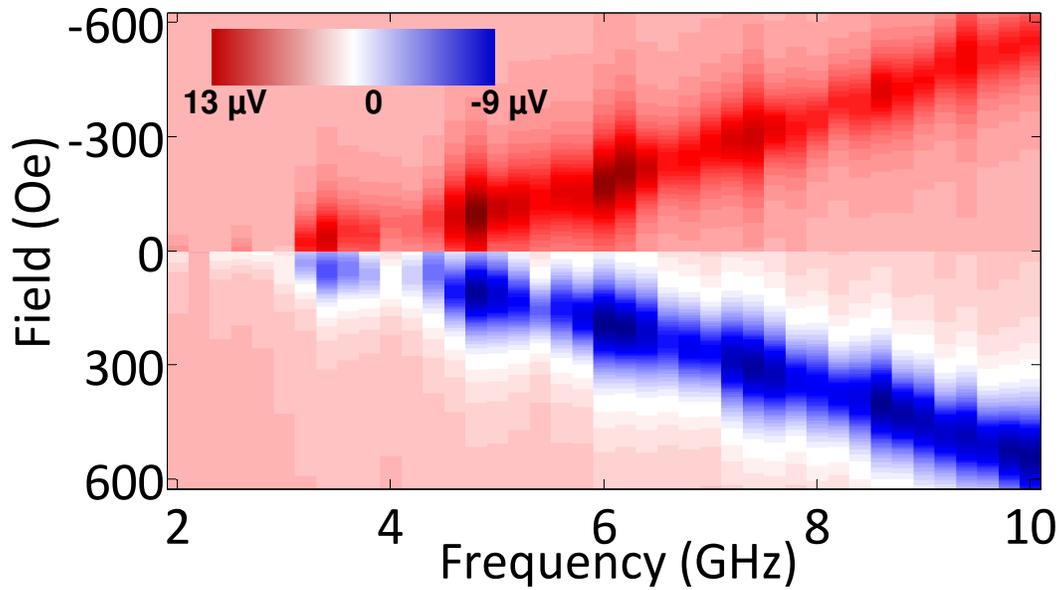

Figure S4: The contour image of the spin pumping spectra at various excitation frequencies for Sample BS2.

## 2. Extraction of magnetic properties

The saturation magnetization (*M*s) and gyromagnetic ration (*γ*) of the magnetic layer are extracted for each of the samples BS1-BS4 by fitting of the Kittel formula $f = \dfrac{\gamma}{2\pi}\sqrt{H(H+M_s)}$ over data in Fig. 5(c) in the main text. Figure S5 shows fitting results overlaid on experimental data of the sample BS1. The resultant values of $M_s$ are 1210, 1220, 1050 and 1370 emu/cc for BS1-BS2, respectively. In addition, the resultant value for *γ* is about $2.7\times10^5$ s$^{-1}$.T$^{-1}$ in all the samples.

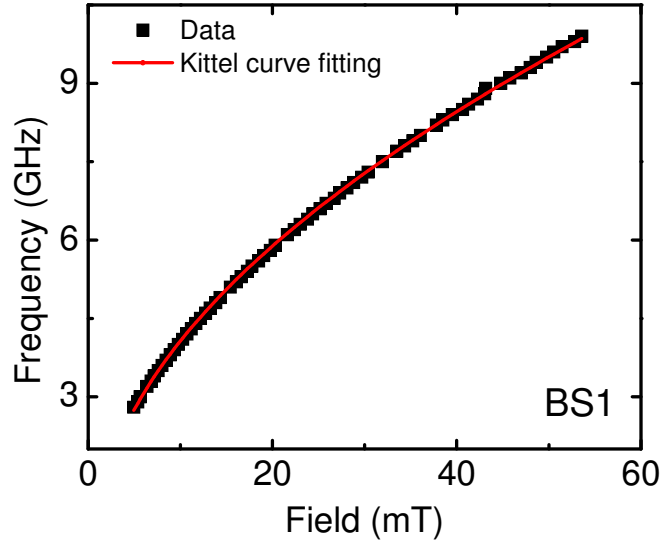

Figure S5: The curve fitting of the Kittel formula over the experimental results from sample BS1

## 3. Characterization of the coplanar waveguide input impedance

In order to calculate the rf-field accurately, we have measured the input impedance of the coplanar waveguide using Anritsu 37397D vector network analyzer (VNA) as shown in Fig. 6(a).

The nominal output power given by $P_{out} = P_{in} \dfrac{Z_L(\omega)}{Z_L(\omega)+50}$ has been measured for different excitation frequencies. The normalized output power is shown in Fig. 6(b) where data is normalized for a perfect 50 Ω impedance load. The rf-field generated by the input microwave signal is corrected at different excitation frequencies according to these results.

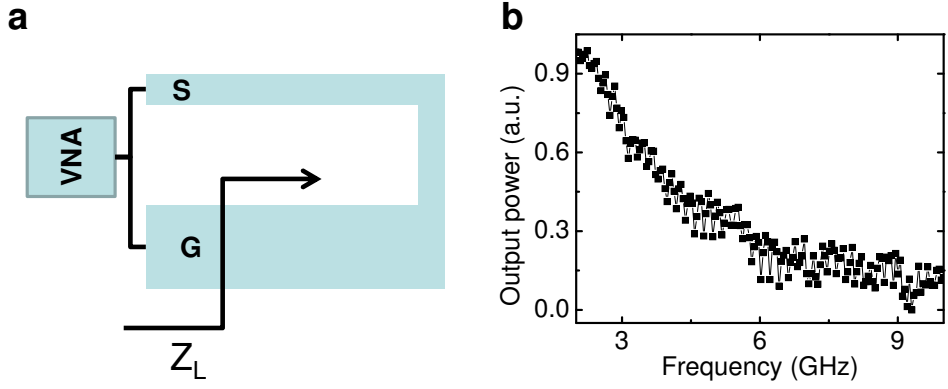

Figure S6: (a) Characterization of the input impedance of the waveguide. (b) The normalized output power of VNA for different excitation frequencies. The data are normalized by a 50 Ω impedance.

4. *Extraction of the spin Hall angle*

As discussed in the main text, the spin current injected into the Bi$_2$Se$_3$ layer can be calculated from $j_s = \dfrac{g^{\uparrow\downarrow}\gamma^2 h_{rf}^2 \hbar [M_s\gamma + \sqrt{(M_s\gamma)^2 + 4\omega^2}]}{8\pi\alpha^2[(M_s\gamma)^2 + 4\omega^2]}$. $h_{rf}$ is calculated according to Ampere law $h_{rf} = I_{rf}/(2*W)$ where $W$ is the width of the coplanar waveguide (60 μm) and $I_{rf}$ is the input microwave current which is corrected based on the VNA results in Fig. 6(b). The damping is calculated based on $\alpha = \dfrac{\sqrt{3}\gamma\Delta H}{2\omega}$ and are given for different excitation frequencies in Fig. S7.

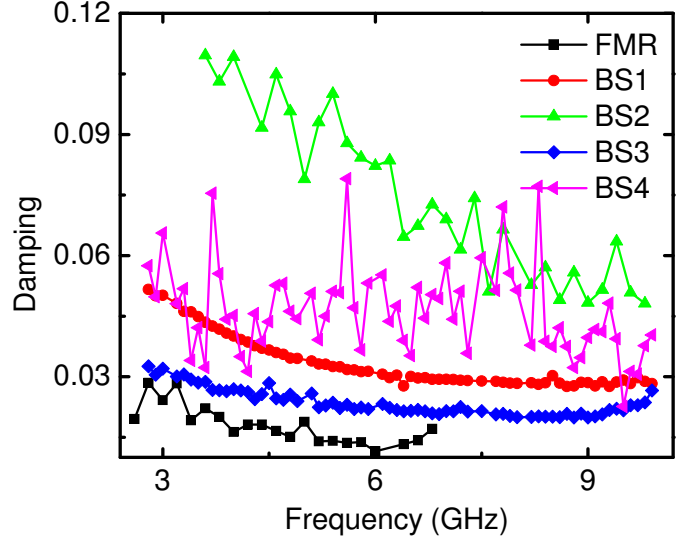

Figure S7: The damping value of the sample BS1-BS4 for different excitation frequencies. The damping of the FMR sample is also given for reference.

Based on the output voltage, the spin Hall angle can be calculated from $V_{ISHE} = \frac{w \theta_{SHE} \lambda_N \tanh(d_N/2\lambda_N)}{d_N \sigma_N + d_F \sigma_F}(\frac{2e}{\hbar})j_s$. $\lambda_N$ is the spin coherence length and is assumed to be about 5 nm which is inline with other report[4]. The resistivity of CoFeB is measured to be about 84 μΩ.cm. The width of the CoFeB that contributes to the output voltage is $w$. Considering the spin wave contribution, it is 60 μm + 2*$L_{SW}$ + (180 μm + 2*$L_{SW}$)/3[4]. The spin wave propagation length for the CoFeB is about 25 μm[5] while present of $Bi_2Se_3$ at the interface with CoFeB damped the spin wave. The spin wave propagation length is calculated from $L_{SW}$=(25 μm)*$\alpha_{FMR}/\alpha_{SP}$. Substitution of these parameters in above equation gives an estimate of the spin Hall angle.